\documentclass[9pt,twocolumn,twoside]{pnas-new}

\usepackage{fancyhdr}

\templatetype{pnasresearcharticle} 

\title{Viscosity Metamaterials}

\author[a,1,2]{Prateek Sehgal}
\author[b,2]{Meera Ramaswamy} 
\author[b]{Edward Y. X. Ong}
\author[c]{Christopher Ness}
\author[b]{Itai Cohen}
\author[a,d]{Brian J. Kirby}

\affil[a]{Sibley School of Mechanical and Aerospace Engineering, Cornell University, Ithaca, New York 14853, USA}
\affil[b]{Department of Physics, Cornell University, Ithaca, New York 14853, USA}
\affil[c]{School of Engineering, University of Edinburgh, Edinburgh EH9 3FG, United Kingdom.}
\affil[d]{Division of Medicine, Division of Hematology and Medical Oncology, Weill--Cornell Medicine, New York, New York 10021,USA}

\leadauthor{Sehgal} 

\authorcontributions{PS, MR, IC, and BJK conceived of experiments. PS, MR, EYXO performed experiments. PS and MR analyzed data. CN performed simulations. PS, MR, IC, and BJK wrote the manuscript. }
\authordeclaration{Authors declare no competing interests.}
\correspondingauthor{\textsuperscript{1} To whom correspondence should be addressed. E-mail: ps824@cornell.edu. \newline \textsuperscript{2} These authors contributed equally to this work.}


\begin{abstract}
Metamaterials are composite structures whose properties arise from a mesoscale organization of their constituents. Provided this organization occurs on scales smaller than the characteristic lengths associated with their response, it is often possible to design such materials to have properties that are otherwise impossible to achieve with conventional materials-- including negative indexes of refraction, perfect absorption of electromagnetic radiation, and negative Poisson ratios \cite{StructuralMeta, OpticalMeta, ThermalMeta_Cloaking, AbsorbingMetamaterial, Brunet2015, HydrodynamicMetaCloak, park2019fluid, park2021metamaterial, AcousticMeta}. Here, we introduce and demonstrate a new material class-- viscosity metamaterials. Specifically, we show that we are able to rapidly drive large viscosity oscillations in a shear-thickened fluid using acoustic perturbations with kHz to MHz frequencies. Because the time scale for these oscillations can be orders of magnitude smaller than the timescales associated with the global material flow, we can construct metamaterials whose resulting viscosity is a composite of the thickened, high-viscosity and dethickened, low viscosity states. Such viscosity metamaterials can be used to engineer a variety of surprising properties including negative viscosities, a response that is inconceivable with conventional fluids. The high degree of control over the resulting viscosity, the ease with which they can be accessed, and the variety of exotic properties achievable by viscosity metamaterials make them attractive for uses in technologies for which control over fluid flows and their instabilities are critical, ranging from coatings to cloaking to 3D printing.

\end{abstract}

\dates{This manuscript was compiled on \today}
\setboolean{displaywatermark}{false}

\begin{document}

\maketitle
\ifthenelse{\boolean{shortarticle}}{\ifthenelse{\boolean{singlecolumn}}{\abscontentformatted}{\abscontent}}{}

\begin{figure*}[t]
\centering
\includegraphics[width=0.8\linewidth]{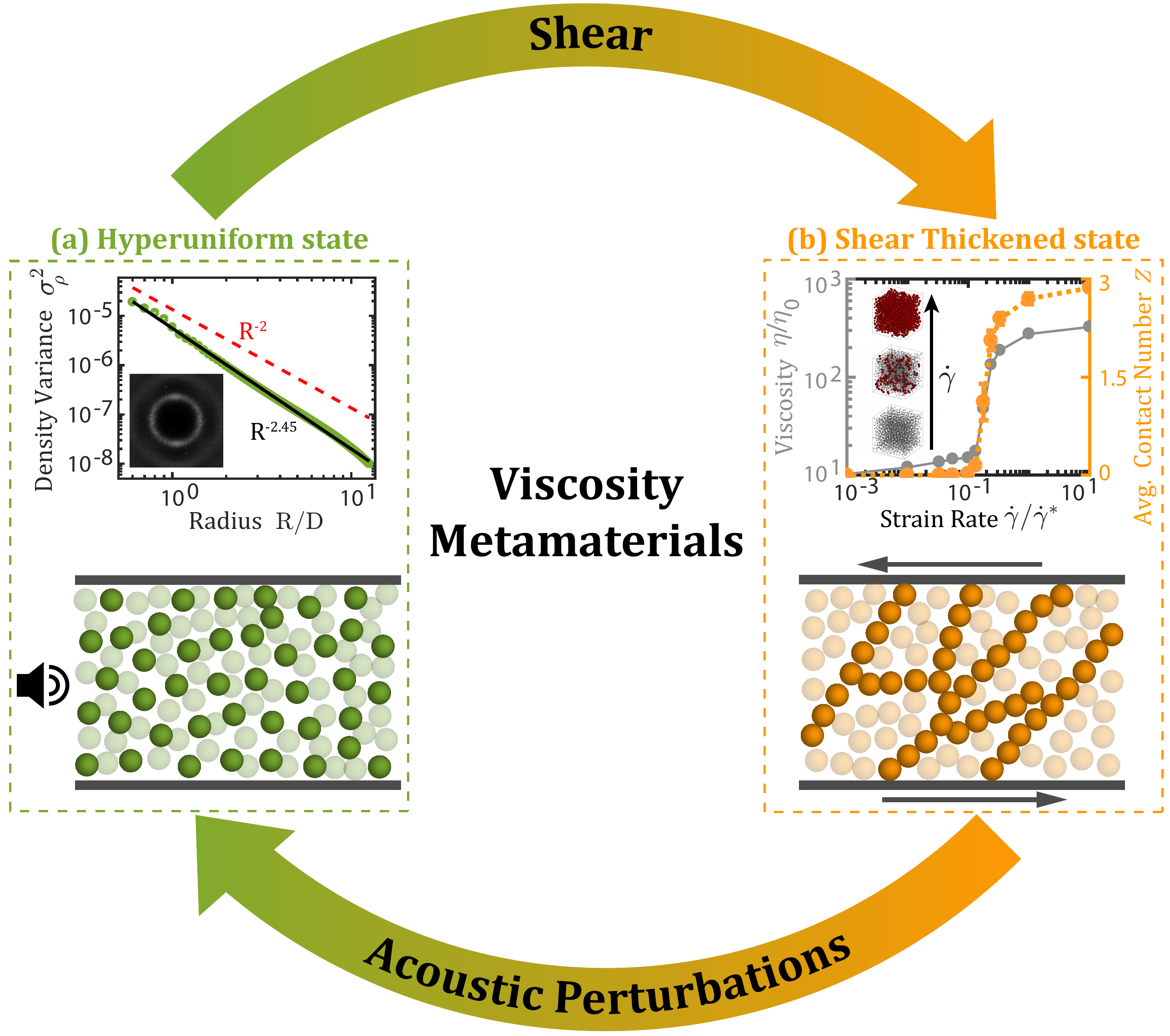}
\caption{Viscosity metamaterials generated by the interplay of acoustic perturbations and applied shear. (a) (Top) Evolution of number density variance $\sigma_\rho^2$ (green circles) with radius $R/D$ of the observation window in the presence of acoustic perturbations. Here, $D$ is the particle diameter and red dashed line shows $R^{-2}$ scaling of a completely random system for reference. $\sigma_\rho^2$  decays faster than $R^{-2}$ and approximately scales with $R^{-2.45}$ (solid black line), which is similar to the exponent observed for hyperuniform absorbing states at criticality \cite{Hexner2015,Wilken2020}. (Inset) The spectrum of structure factor $S(\textbf{k})$, with $\textbf{k}=0$ at the center. Forward scattering, i.e., $S(\textbf{k}=0)$, is omitted from the spectrum. $S(\textbf{k})$ exhibits a dark region in the center (at small $k$) surrounded by a bright ring. The behavior of both $\sigma_\rho^2$ and $S(\textbf{k})$ are hallmarks of disordered hyperuniform states \cite{Torquato2003,Zachary2009,Dreyfus2015,Torquato2018}. The data is obtained from the average of over 1200 images from the timelapse. (Bottom) Schematic representation of the hyperuniform arrangement of particles in the presence of acoustic perturbations. (b) (Top) Flow curve showing suspension viscosity $\eta/\eta_0$ (left axis) and contact number Z (right axis) as functions of shear rate $\dot\gamma/\dot\gamma^*$ obtained from particle-based simulation (see Materials and Methods). (Inset) Snapshots of the  suspension microstructures at increasing shear rate. Particles in red have at least one contacting particle while those in grey have none. We observe formation of the force chains with increasing shear rate. (Bottom) Schematic representation of the force chain network in a shear-thickened state. (Large circular arrows) Toggling between the hyperuniform state and shear thickened state allows for generating viscosity metamaterials.}
\label{fig:figure1}
\end{figure*}

Our inspiration for designing viscosity metamaterials arises from the tunability of a shear-thickened suspension's viscosity via acoustic perturbations \cite{Sehgal2019}. Under shear, nearly all dense suspensions undergo a thickening transition from a low-viscosity to a high-viscosity state \cite{Denn2018, Wagner_PhysTod,BrownandJaeger2014,Barnes_review}. This transition occurs when the applied shear stresses are large enough to drive particles into contact, forming force chains that are primarily aligned along the maximum compressive axis \cite{Lin_dethick,MariandSeto_ForceChains,Cates_ForceChains,Majmudar_ForceChains,denn2014rheology,BrownJaeger2012,Wyart_DSTFriction,Fernandez_FricTransition,Royer_Friction, Singh2020, Lin_ShearReversal,Bi_jamming,Morris_Review,SetoandMari_FrictionDST, Ness_CrossShear}. The resulting increase in viscosity can span orders of magnitude and in sufficiently dense suspensions can result in a completely rigid, shear jammed state. Recently, we have shown that by applying acoustic perturbations the suspension viscosity can be tuned from the thickened to the fully dethickened state \cite{Sehgal2019}. The applied acoustic perturbations are presumed to dethicken the suspension by breaking up particle contacts important for force transduction. Importantly, since acoustic perturbations can be applied at kHz to MHz frequencies, there is an opportunity to generate new material properties via rapidly cycling between the thickened and dethickened states (Fig.~\ref{fig:figure1}). 

We pursue this strategy using charge stabilized monodispersed colloidal suspensions consisting of 2-$\mu$m silica particles suspended in dipropylene glycol ($\eta_0 = 0.11 \text{ Pa s}$) at a volume fraction of $\phi \simeq 0.53$. When we apply acoustic perturbations to a quiescent suspension (for apparatus and analysis, see methods), we find that particles form a disordered hyperuniform state, as shown by the behavior of the number density variance and structure factor in Fig. \ref{fig:figure1}(a). These results indicate that acoustic perturbations generate isotropic repulsion that maximizes the interparticle separations as shown in the schematic at the bottom of Fig.~\ref{fig:figure1}(a). Conversely, as the suspension thickens under shear, particles form contacts, organized into anisotropic force chains that align with the maximum compressive axis. The increase in viscosity arises from the increase in average number of contacts between the particles \cite{pradeep2021jamming, MariandSeto_ForceChains, Morris_Review, Lin_ShearReversal} (Fig.~\ref{fig:figure1}(b)). In simulations (see methods and SI for details), the force chains formed can be visualized by plotting the 3-dimensional microstructure as seen in the inset of Fig.~\ref{fig:figure1}(b). These results illustrate that there is an opportunity for toggling between these two structural arrangements to form viscosity metamaterials.

To investigate whether such oscillations are indeed possible, we develop acoustic protocols and implement them using a previously described acoustic apparatus mounted on an Anton Paar MCR Rheometer \cite{Sehgal2019}. In particular, to generate the oscillations in the viscosity, we apply the acoustic fields intermittently. We anticipate that over the period where the acoustic perturbations are on $T_\mathrm{on}$, the viscosity will rapidly drop due to dissolution of force chains. Conversely, over the period where the perturbations are turned off $T_\mathrm{off}$, formation of force chains is unimpeded, and the viscosity increases.

\begin{figure*}[t]
\centering
\includegraphics[width=0.9\linewidth]{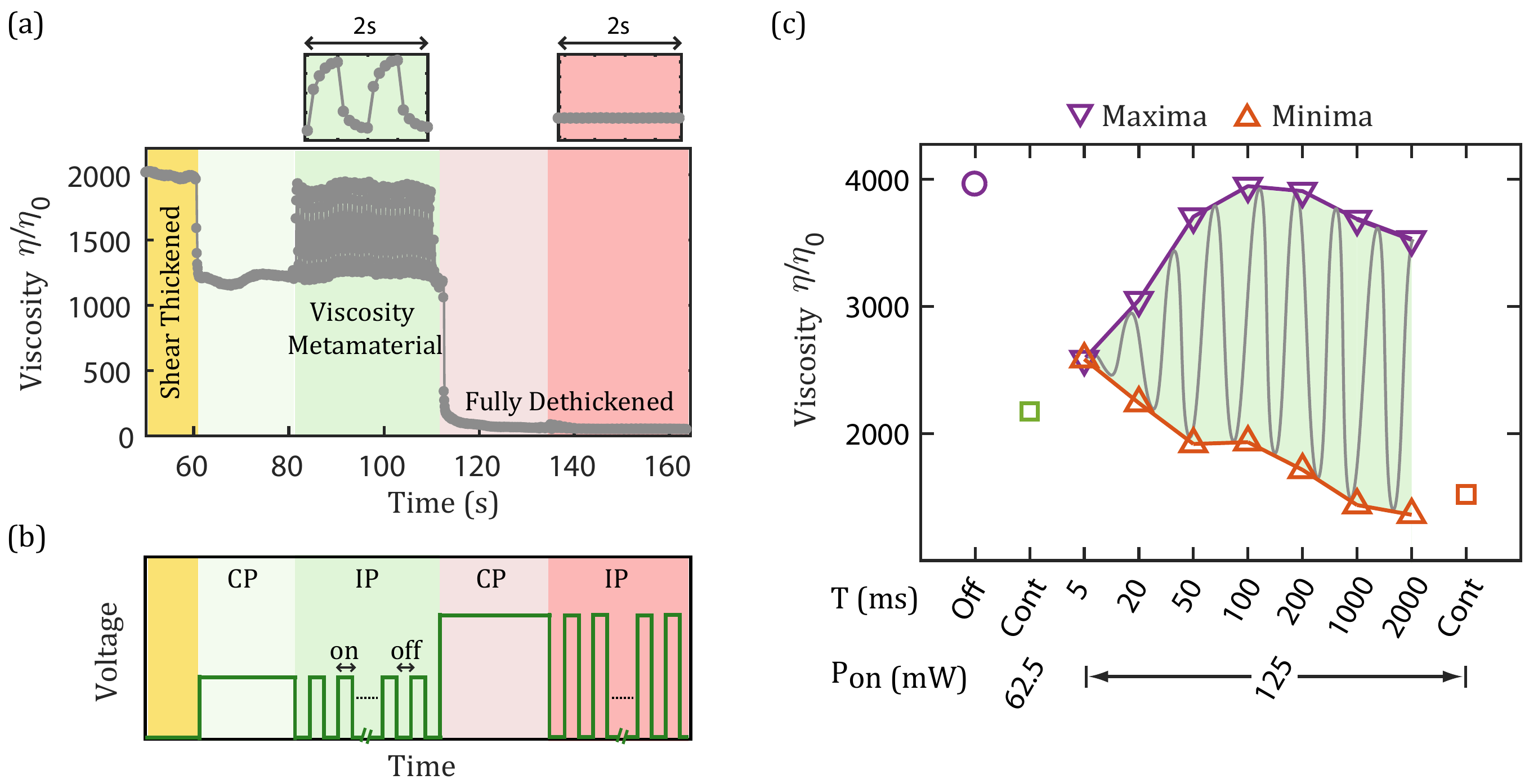}
\caption{Generating viscosity metamaterials by use of shear and intermittent acoustic perturbations. (a) Instantaneous viscosity response of the suspension to the continuous perturbations (CP) and intermittent perturbations (IP). Here, $\eta_0$ is the solvent viscosity (dipropylene glycol, 0.11 Pa s). The measurement is performed at a steady strain rate of $\dot{\gamma} =0.44 \text{ s}^{-1}$. The suspension maintains a shear-thickened state for the first 60 s. After thickening, the suspension is partially dethickened by applying 3~V peak-to-peak ($P_\text{on}=22.5$~mW) CP (light green). Next, the IP are applied with $T_\text{on} = 0.5$~s and $T_\text{off} = 0.5$~s (green), resulting in rapid oscillations of viscosity and formation of a viscosity metamaterial. Following this, the suspension is fully dethickened by applying 10~V peak-to-peak ($P_\text{on}=250$~mW) CP (light orange). When the IP with $T_\text{on} = 0.5$~s and $T_\text{off} = 0.5$~s are applied in this state (orange), the viscosity does not oscillate.  This portion of the phase space does not support a viscosity metamaterial, but simply an acoustically dethickened state (see SI for discussion on fully dethickened states). The zoomed-in views (top) show the viscosity response for two cycles of the IP. (b) Schematic of the input voltage waveform for the acoustic protocol. (c) Tunability of the viscosity metamaterials. The plot shows oscillations of viscosity with the time period $T$ of IP.  Purple and orange triangles show the average maximum and minimum of the oscillatory viscosity during IP. The hand-drawn grey curve illustrates the oscillations of viscosity. The on and off durations of the IP are kept equal and specified by $T_\text{on}=T_\text{off}=T/2$. The power of IP during the on-time ($P_\text{on}$) is set to 125~mW. Green and orange squares show the viscosity level obtained from CP of powers $P_\text{on}=62.5$~mW and $P_\text{on}=125$~mW, respectively. The viscosity oscillations at $T=5$~ms are comparable to the sensitivity of our rheometer, thus we plot the average viscosity instead of the maximum/minimum viscosity for this condition. The measurements are performed at a constant strain rate $\dot{\gamma} =0.57 \text{ s}^{-1}$. }
\label{fig:figure2}
\end{figure*}

The response of the suspension sheared at $\dot\gamma = 0.44$~s$^{-1}$ to our acoustic protocols is shown in Fig.~\ref{fig:figure2}(a),(b). Under steady shear with no applied acoustic perturbations, the suspension maintains a shear-thickened viscosity. Application of continuous acoustic perturbations at 1.15~MHz, 5~V partially dethickens the suspension. Toggling these perturbations on and off with $T_\mathrm{on} = 0.5$~s and $T_\mathrm{off} = 0.5$~s results in rapid oscillations between the fully thickened and a partially dethickened state, indicating the formation of a viscosity metamaterial.

To illustrate the remarkable tunability of these materials, we generate metamaterial states with roughly the same average, but vastly different ranges of maximum and minimum viscosities (Fig.~\ref{fig:figure2}(c)). Specifically, we vary the time period of the intermittent perturbations (IP), keeping $T_{\text{on}} = T_{\text{off}}$, and plot the envelope of the viscosity oscillations with the maximum indicated by purple inverted triangle and minimum indicated by the orange triangles. When the period of the acoustic perturbations is short (5~ms), the amplitude of the viscosity oscillations (illustrated schematically by the grey curve) is small. Here, the minimum and the maximum viscosity remain close to the value obtained when continuous acoustic perturbations are applied at half the power (green square). This behavior suggests that a time scale of 5~ms is too short for force chains to evolve at this shear rate.\footnote{Although  at low $\mathrm{Re}$, the \textit{fluid physics} has no inherent timescale, there exist strain scales in the system corresponding to force chain formation or breakup. We refer informally to the time scale of the \textit{system} as the ratio of the strain scale to the strain rate.  This enables comparisons between the temporal manifestation of the strain scale and the time durations applied via acoustics.} As the period of acoustic perturbations is increased, the amplitude of the viscosity oscillations increases dramatically, with a viscosity maximum that approaches the fully thickened state (purple circle) and minimum that is comparable to the lowest viscosity attainable at this applied acoustic power (orange square). Thus, by varying the acoustic power as well as the periods $T_\mathrm{on}$ and $T_\mathrm{off}$, one can easily generate oscillations over the entire range of viscosities associated with thickening, all while the fluid is continuously sheared.

\begin{figure*}[t]
\centering
\includegraphics[width=\linewidth]{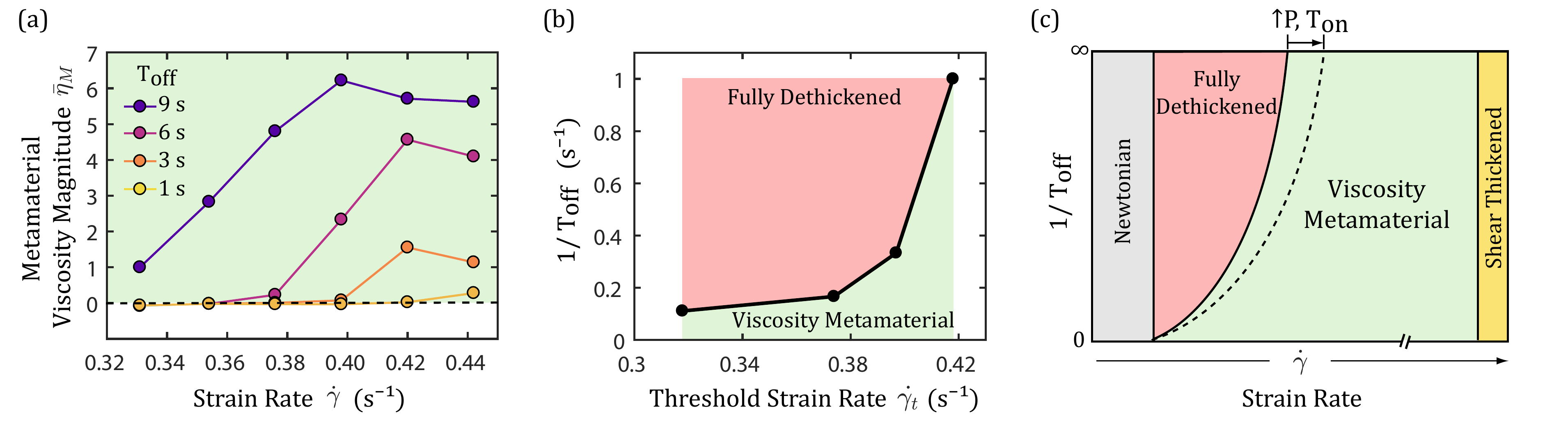}
\caption{Parameter space governing the formation of viscosity metamaterials. (a)  Metamaterial viscosity magnitude $\bar{\eta}_{M}$, defined as normalized difference between the time-averaged viscosities under intermittent perturbations and continuous perturbations $(\bar{\eta}_{IP}-\bar{\eta}_{CP})/{\bar{\eta}_{CP}}$, as a function of strain rate $\dot{\gamma}$ and off times $T_{\text{off}}$.  The viscosity metamaterial region is depicted by the green color and the fully dethickened state is depicted by the dashed black line. These measurements are performed at $T_\text{on} = 1~\text{s}$ and $P_\text{on} = 250~\text{mW}$. (b) Phase boundary between the fully dethickened and viscosity metamaterial regions. The threshold strain rate $\dot{\gamma}_t$ is defined as the maximum strain rate at a given $T_\text{off}$ for which $\bar{\eta}_{M}\simeq 0$. (c) Phase diagram illustrating the viscosity metamaterial region in $1/T_{\text{off}}$ versus $\dot{\gamma}$ parameter space. The diagram is constructed for a constant $P_{\text{on}}$ and $T_{\text{on}}$. The dashed line indicates how the phase boundary between the fully dethickened and viscosity metamaterial regions shifts as dethickening energy increases because of increase in power ($P_{\text{on}}$) or duration ($T_{\text{on}}$) of the acoustics.}
\label{fig:figure4}
\end{figure*}

To illustrate the broad range of parameters over which acoustically generated viscosity metamaterials can be accessed, we determine the threshold strain rates and periods $T_\mathrm{off}$, beyond which it is possible to attain metamaterial states. Crucially, to obtain metamaterial states, the strain rate must be sufficiently high for the suspension to thicken via formation of force chains during the period where the acoustics are off, $T_\mathrm{off}$. For example, a strain rate of $\dot{\gamma} =0.44 \text{ s}^{-1}$ is insufficient to produce viscosity oscillations when the suspension is fully dethickened and intermittent perturbations with $T_{\text{off}} = 0.5$~s are applied (Figure~\ref{fig:figure2}(a) orange shading; see SI for details on fully dethickened states). To determine the threshold strain rate for each $T_{\text{off}}$, we note that in the metamaterial state the time-averaged viscosity under continuous acoustic perturbations, $\bar{\eta}_{CP}$, is always smaller than the time-averaged viscosity under intermittent acoustic perturbations, $\bar{\eta}_{IP}$, whereas in the fully dethickened state, these viscosities are identical. Thus, for each value of $T_{\text{off}}$ we measure the magnitude of metamaterial viscosity $\bar{\eta}_{M}$ defined as the normalized difference of the time averaged viscosities ($\bar{\eta}_{IP}-\bar{\eta}_{CP})/{\bar{\eta}_{CP}}$ (see experimental protocols in supplementary materials) and determine for $T_\mathrm{on} = 1~\text{s}$ and $P_\mathrm{on} = 250~\text{mW} $ the threshold strain rate beyond which it becomes greater than zero (Fig.~\ref{fig:figure4}(a)). As expected, we find that as $T_{\text{off}}$ decreases, a larger strain rate is necessary to generate the metamaterial state (Fig.~\ref{fig:figure4}(b)).


Collectively, from these observations, we composed a qualitative phase diagram describing the region in the $1/T_{\text{off}}$ versus $\dot{\gamma}$ parameter space where viscosity metamaterials can be generated (Fig.~\ref{fig:figure4}(c)). At low shear rates, the viscosity is Newtonian (grey) and insensitive to acoustic perturbations. Similarly, we anticipate that at extremely high shear rates inaccessible in our rheology measurements, the applied power will be insufficient to dethicken the system and the viscosity reflects the fully thickened state (yellow). At intermediate shear rates and finite values of $T_{\text{off}}$, the boundary between the fully dethickened (orange) and viscosity metamaterial region (green) is governed by the interplay between acoustic perturbations and shear flows as measured in Fig.~\ref{fig:figure4}(b). Moreover, these boundaries can also be shifted with the acoustic power $P_{\text{on}}$ and $T_{\text{on}}$ (dashed lines). Importantly, at each point within the viscosity metamaterial region, we can dramatically tune the viscosity envelope that characterizes the metamaterial (Fig.~\ref{fig:figure2}(c)). Specifically, by precisely controlling $T_{\text{off}}$, $T_{\text{on}}$, the acoustic power, and strain rate, vastly different combinations of the minimum, average, and maximum viscosity can be obtained. As such, acoustic perturbations enable us to access novel metamaterial states over a broad range of parameters, with the potential to engineer responses that are impossible to achieve in conventional fluids.



\begin{figure*}
\centering
\includegraphics[width=\linewidth]{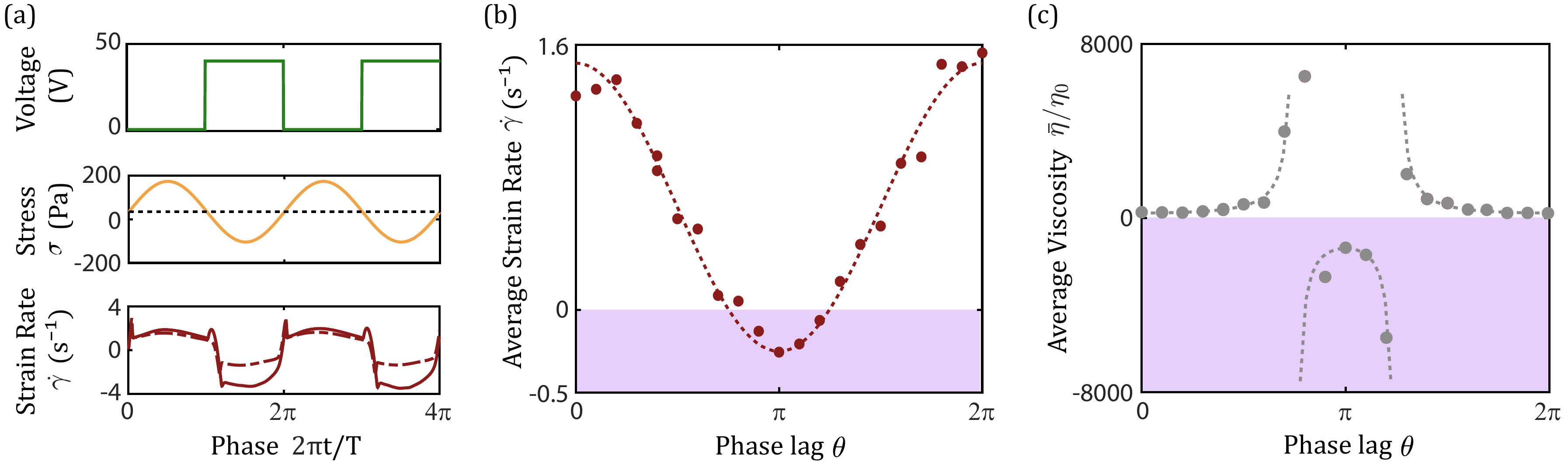}
\caption{Negative-viscosity states can be achieved in oscillatory systems by controlling the phase relation between shear and voltage. (a) The applied voltage (top) and stress (middle) and the resultant strain rate (bottom). An oscillatory stress (orange) with a positive bias (dotted line) is used to drive the sample. Bottom panel indicates the resultant strain rate in the absence (dashed line) or presence (solid line) of acoustics.  (b) The time-averaged strain rate for various phase lags between the acoustics and applied stress. When acoustics is applied with a phase lag $\theta$, the resultant average strain rate is non monotonic, and is well described by a $\dot\gamma=a(1+b\cos\,\theta)$ dependence (dashed line). Interestingly, the strain rates are negative at phase lags close to $\pi$ (purple region) (c) Time-averaged viscosities normalized by the solvent viscosity for various applied phase lags. The averaged viscosity $\bar{\eta}/\eta_0$ is well described by $(1+b\cos\,\theta)^{-1}$ functional form (dashed line), which generates negative viscosity (denoted in purple) for phase lags near $\pi$.} 
\label{fig:figure5}
\end{figure*}

To illustrate one such response, we use viscosity metamaterials to engineer a negative viscosity. 
Such negative viscosity states have been reported in previous work on active bacterial suspensions \cite{Loisy2018_NegVisc}, and in shear thickening fluids driven by oscillatory shear \cite{hou2021study}. In contrast to the viscosity metamaterials described here, however, such responses could only be achieved over an extremely limited range of parameters. To achieve negative viscosities in our system, we oscillate the driving stress sinusoidally with a positive bias and apply acoustics to generate low viscosity states during selective points in the cycle (Fig.~\ref{fig:figure5}(a)). If the acoustics are applied in phase with the stress ($\theta = 0$), the low viscosity states are generated while the suspension is flowing forward, resulting in an effective low positive viscosity state. In contrast if the acoustics are applied with a phase lag $\theta = \pi$, the low viscosity states are generated while the suspension is flowing \textit{backwards}, resulting in a net negative strain rate (Fig.~\ref{fig:figure5}(b)). In this scenario, the suspension flows in the direction opposite to the average applied stress, and displays an negative effective viscosity. 



The full range of behaviours demonstrated by the average strain rate and the effective viscosities at different phase lags is shown in Fig.~\ref{fig:figure5}(c). In addition to the negative viscosity states observed when the phase difference is close to $\pi$, we can also find values of $\theta$ where the strain rate is zero, indicating the formation of states with infinite viscosity, where the suspension remains stationary, despite experiencing a net positive stress. Similar protocols can also be employed to engineer zero viscosity systems (See SI). We note that to achieve these exotic states we are pumping energy into the system via the acoustics and are therefore not violating any thermodynamic laws. These results offer just a glimpse of the exotic behaviours displayed by viscosity metamaterials achievable by simple manipulation of the shear and acoustic forces.  

Because these precise acoustic oscillation protocols requires only that we attach a piezo at the boundary, achieving the demonstrated metamaterial states should be straightforward even in complex flow geometries.
Moreover, similar viscosity oscillations could be implemented in more specialized suspensions whose thickening can be controlled magneto or electro-rheologically \cite{Dong_prespective_ER_2019,morillas2020magnetorheology,eshaghi2016dynamic}. As such, viscosity metamaterials open uncharted avenues for conveniently generating novel fluid flows, transient flow behaviours, and geometry-specific responses, all of which could be used in next-generation applications. These include new methods for processing dense coatings, generating hydrodynamic cloaking \cite{HydrodynamicMetaCloak,Urzhumov_2011,Culver2017,park2020assembling,pang2022hydrodynamic,wang2021intangible, boyko2021microscale}, and processing dense suspension in microfluidic environments during 3D printing~\cite{ONeillExtrusion2019}.

\subsection*{Supporting Information Appendix (SI)}

\matmethods{\textbf{Apparatus:} We work with charge-stabilized, monodispersed colloidal suspensions consisting of 2-$\mu$m silica particles suspended in dipropylene glycol ($\eta = 0.11 \text{ Pa s}$) at a volume fraction of $\phi \simeq 0.53$. The test setup consists of a piezoelectric disk bonded via epoxy to a custom-made aluminum bottom-plate (details in \cite{Sehgal2019}). The acoustic perturbations are generated by applying an AC voltage signal to the piezoelectric disk at resonance frequency $f=1.15 \text{ MHz}$ corresponding to its thickness mode of vibration. For imaging the particle distributions in the presence of perturbations, the piezo-plate setup is integrated with a Zeiss 5 Live inverted confocal microscope, wherein the suspension is confined between the aluminum bottom-plate and a glass slide. For measuring the rheological response in the presence of acoustic perturbations and shear, the piezo-plate setup is integrated with the Anton-Paar MCR 702 Rheometer. The glass top plate is used to apply steady shear to the suspension and measure the shear viscosity. The gap between the bottom plate and the top plate is set to 0.64 mm.

\textbf{Simulations:}
We simulated the trajectories of $N=2000$ spheres with radii $a$ and $1.4a$ in a periodic box with length $L$ chosen so that the volume fraction $\phi \equiv(2\pi/3)Na^3(1 + 1.4^3)/L^3 = 0.55$. Particles are subjected to hydrodynamic forces $\mathbf{F}_h =\mathcal{R}\mathbf{V}$ (where $\mathcal{R}$ is a resistance matrix comprising Stokes drag and pairwise lubrication terms); repulsive forces $\mathbf{F}_r=F_{r,0}\exp(-h\kappa)$ (with $h$ the dimensionless surface-surface separation and $\kappa\approx10^2$); and frictional contact forces $\mathbf{F}_c$ activated only when $h<0$, and their trajectories are updated using a velocity Verlet scheme. We shear the simulation box at constant rate, and the suspension viscosity $\eta/\eta_0$ is controlled by $\dot{\gamma}/\dot{\gamma}^* = \dot{\gamma}\eta_0a^2/F_{r,0}$. Further details are given in~\cite{cheal2018rheology}.

\textbf{Protocol for measuring the magnitude of metamaterial viscosity (Fig. ~\ref{fig:figure4}a):} First, we preshear the suspension to achieve the steady state. Following preshear, we shear the suspension at a fixed strain rate for $t\simeq~60$~s. Concurrently, we apply continuous perturbations ($P_\text{c}=250$~mW) for the first 20~s and then apply intermittent perturbations ($P_\text{on}=250$~mW) of desired off-time for the next 40~s. We repeat this process for different strain rates and off times in a systematic manner, while keeping the $P_{\text{on}}$ and $T_{\text{on}}$ fixed. From the instantaneous viscosity response at each condition, we measure the time averaged viscosities during the IP and CP, and calculate the magnitude of metamaterial viscosity as $\bar{\eta}_{M}=(\bar{\eta}_{IP}-\bar{\eta}_{CP})/\bar{\eta}_{CP}$.

\textbf{Protocol for generating the negative-viscosity states (Fig. ~\ref{fig:figure5}):} First, we preshear the suspension to achieve the steady state. Following the preshear, we measure the viscosity at different stresses to identify the optimal mean stress, $\sigma_0 = 34.8$ Pa and the amplitude of the stress oscillations, $\sigma_a = 139.1$Pa. We then apply a stress, $\sigma = \sigma_0 + \sigma_a \sin (2\pi t/T) $, where t is the time in seconds, and $T = 10s$ is the time period of the oscillations. Concurrently, we apply a square acoustic wave by simply turning the piezoelectric disk on and off at the appropriate points on the cycle. We repeat this process over a range of phase lags between the acoustics and the driving stress. We apply two stress oscillation cycles and average the stress, strain rate and viscosity over both the cycles.

}

\showmatmethods{} 

\acknow{The authors would like to thank the Anton Paar VIP academic research program for providing the MCR 702 rheometer, and the Kirby group and Cohen group for valuable discussions. This work is supported by NSF CBET Grants No. 1804963, No. 1232666, and No. 1509308. C.N. acknowledges support from the Royal Academy of Engineering under the Research Fellowship scheme.}

\showacknow{} 

\bibliography{pnas-sample}

\end{document}


\preprint{APS/123-QED}

\title{Viscosity Metamaterials}
\renewcommand{\theequation}{S\arabic{equation}}
\renewcommand{\thefigure}{S\arabic{figure}}
\renewcommand{\thesection}{S\arabic{section}}

\begin{center}
   \large{\textbf{Viscosity Metamaterials}}
    
    \textbf{Supplementary Materials}
\end{center}

\section{Structure factor in simulations}

\begin{figure}[b]
\includegraphics{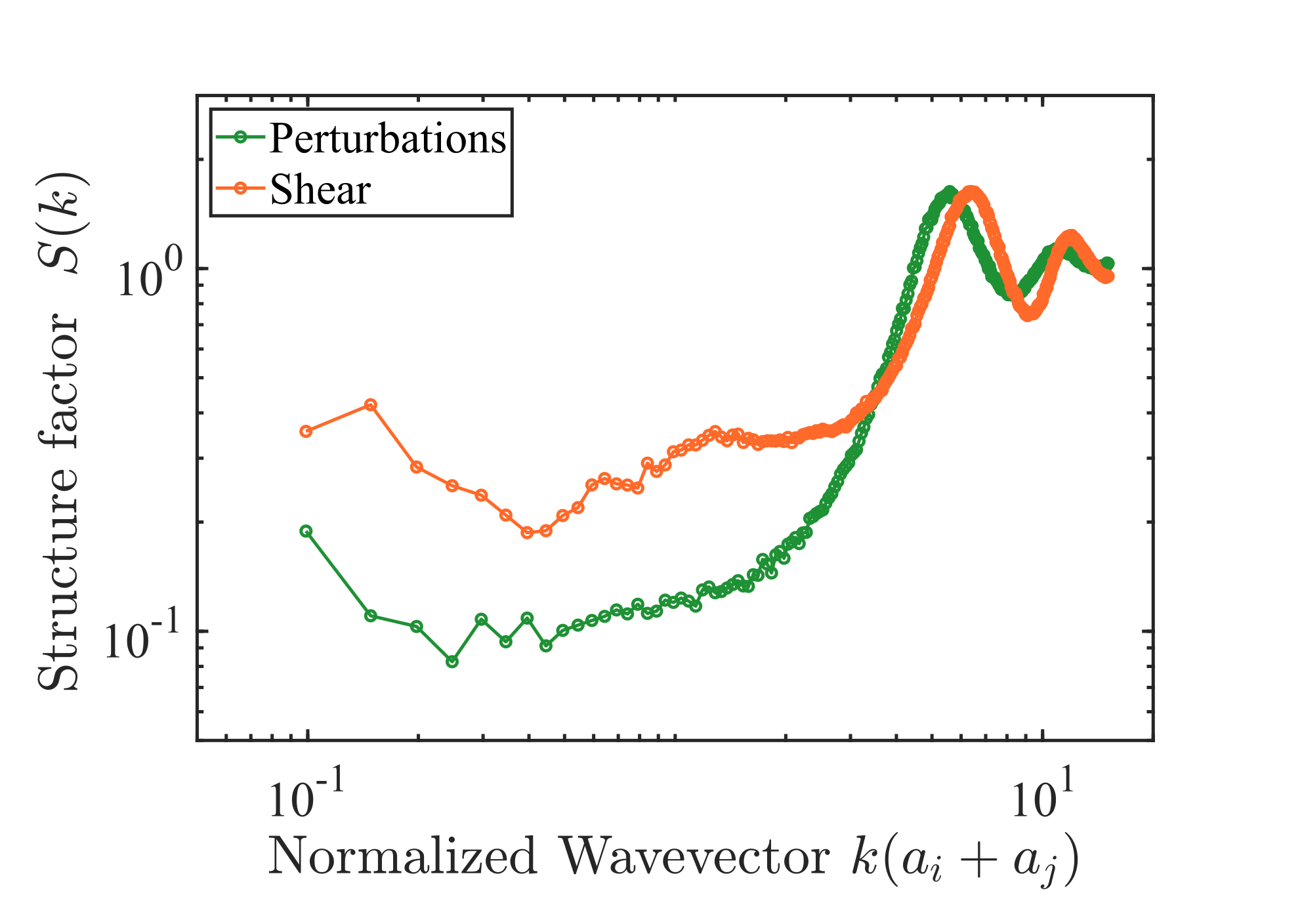}
\caption{\textbf{Structure factor from simulations.} The two distinct structure factors obtains for simulations of 2D discs under shear (orange line) and perturbed via random kick (green line). We observe behaviour consistent with hyperuniformity ($S(k)$ is smaller as $k\rightarrow 0$) when the suspension is perturbed by random kicks.}
\label{fig:S_K_Sim}
\end{figure}

Our ability to generate viscosity metamaterial states indicates that acoustics and shear result in vastly different microstructures, as illustrated by Fig. 1 in the main text. In further support of this idea, we show here the distinct structure factors, $S(k)$ for suspensions under shear and random perturbations obtained via simulations Fig.~\ref{fig:S_K_Sim}.  The simulations were performed in two dimensions using the same framework discussed in the main text. We use bidisperse discs with diameter ratios 1:1.4 to prevent crystallization. The sample is first sheared to steady state, and the resultant structure factor (calculated as in \cite{ness2020absorbing}) is shown in Fig.~\ref{fig:S_K_Sim} (orange line).  Upon attaining steady state, the shear rate is set to zero, and the particles are given random kicks, $\Vec{\psi}$, where the elements of $\Vec{\psi}$ are gaussian with zero mean, and uncorrelated in time. 
The steady state $S(k)$ for the suspension under these random perturbations is shown by the green line in Fig.~\ref{fig:S_K_Sim}. We clearly observe different microstructures because of shear and the random perturbations. Intriguingly, like the formation a hyperuniform state as a result of acoustic perturbations in experimental systems, random perturbations in the simulation systems also appear to drive the suspension towards hyperuniformity as evidenced by a lower $S(k)$ as compared with sheared suspensions as $k \rightarrow 0$. While this decrease in $S(k)$ is not a stringent test for hyperuniformity, it clearly demonstrates that long wavelength density fluctuations are strongly suppressed by the addition of random forces.

\section{Latency in Fully Dethickened States}

We find that the viscosity oscillations can be suppressed in fully dethickened states when the perturbations are applied intermittently  (Fig.~2a orange shading). Specifically, in a fully dethickened state, the viscosity may not increase immediately after the perturbations are turned off. This latency occurs when acoustic perturbations dominate the microstructure of the suspension. We anticipate that the hyperuniform arrangement of particles generated by the perturbations create sufficient separation between the neighbors, leading to a time lag in re-formation of force chains from a fully disrupted network. Importantly, this latency must be overcome by the applied shear in order to generate viscosity metamaterials from a fully dethickened suspension.

\begin{figure}[t]
\includegraphics [width = 0.50\linewidth] {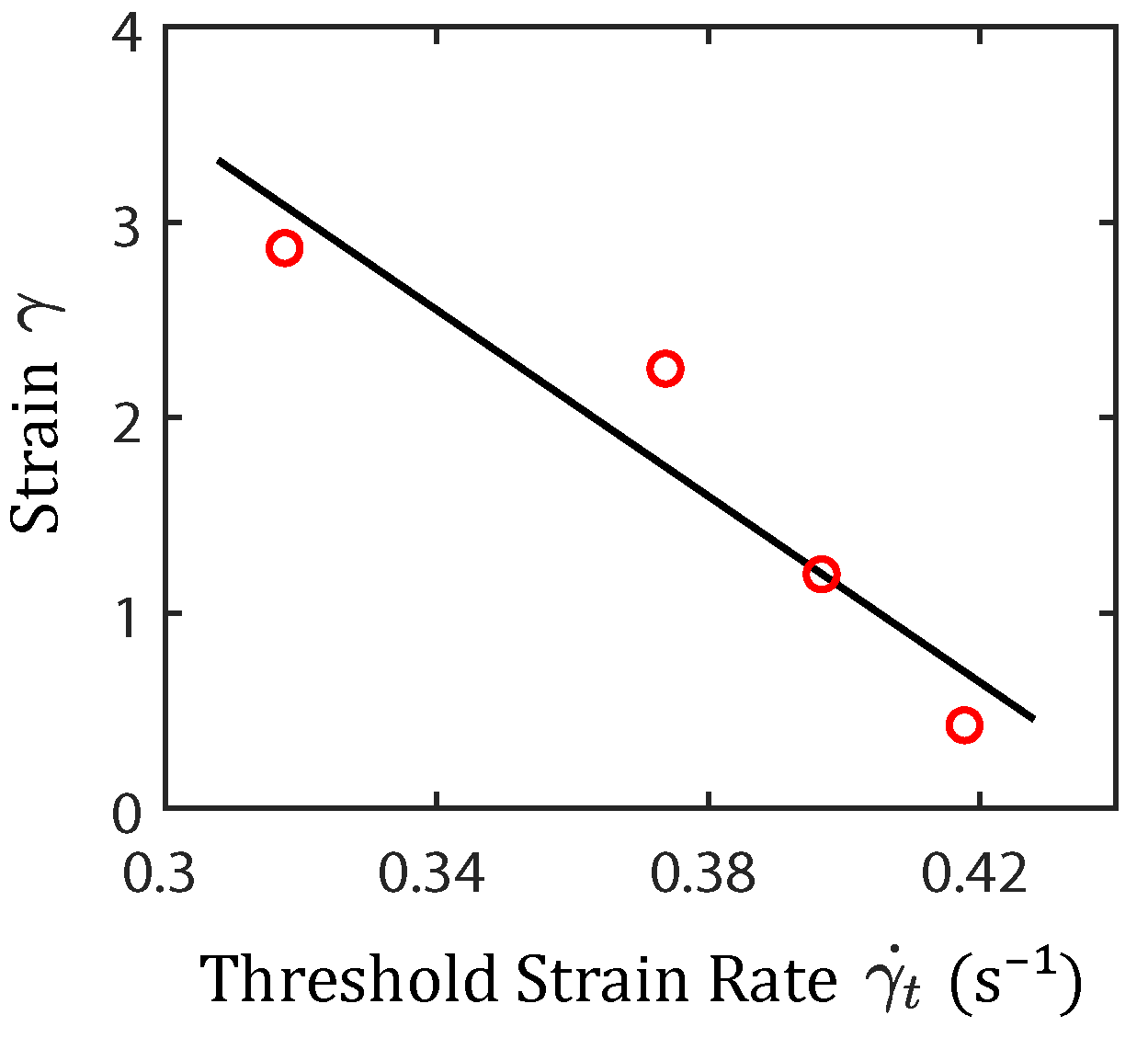}
\caption{\textbf{Strain required to form force chains from a fully dethickened state as a function of strain rate.} Accumulated strain $\gamma$ is calculated from Fig. 3b as $\gamma=\dot{\gamma}_t T_\textrm{off}$. Threshold strain rate $\dot{\gamma}_t$ is same as that in Fig. 3b. Solid black line is a linear fit to the data, presented as a guide to the eye.}
\label{fig:Strain_StrainRate}
\end{figure}

Furthermore, we observed that the strain required to form the force chains from a fully dethickened state decreases with the increasing strain rate (Fig.~\ref{fig:Strain_StrainRate}). This behaviour is in contrast to suspensions under shear alone where reversal experiments have indicated that a fixed strain is required for the formation of force chains \cite{Lin_ShearReversal, ong2020stress}. We argue that this decrease in the required strain with the increasing strain rate also arises from the hyperuniform arrangement of particles generated by acoustic perturbations. Because the perturbations work against the applied shear to keep the particles apart in a hyperuniform manner, we expect the particles to be arranged farther apart from each other at lower strain rates than at higher strain rates. Consequently, a higher amount of strain is required to form the force chains from a fully dethickened state at lower strain rate than that at higher strain rate. From practical standpoint, this collective understanding of the latency and strain dependence in the fully dethickened states can also be leveraged to design smart dethickening protocols for complex flow geometries, in which dethickening is desired at a location where attaching a piezoelectric element is not possible.

\section{Negative Viscosity States}

\begin{figure}[h]
\includegraphics[width = 0.95\linewidth]{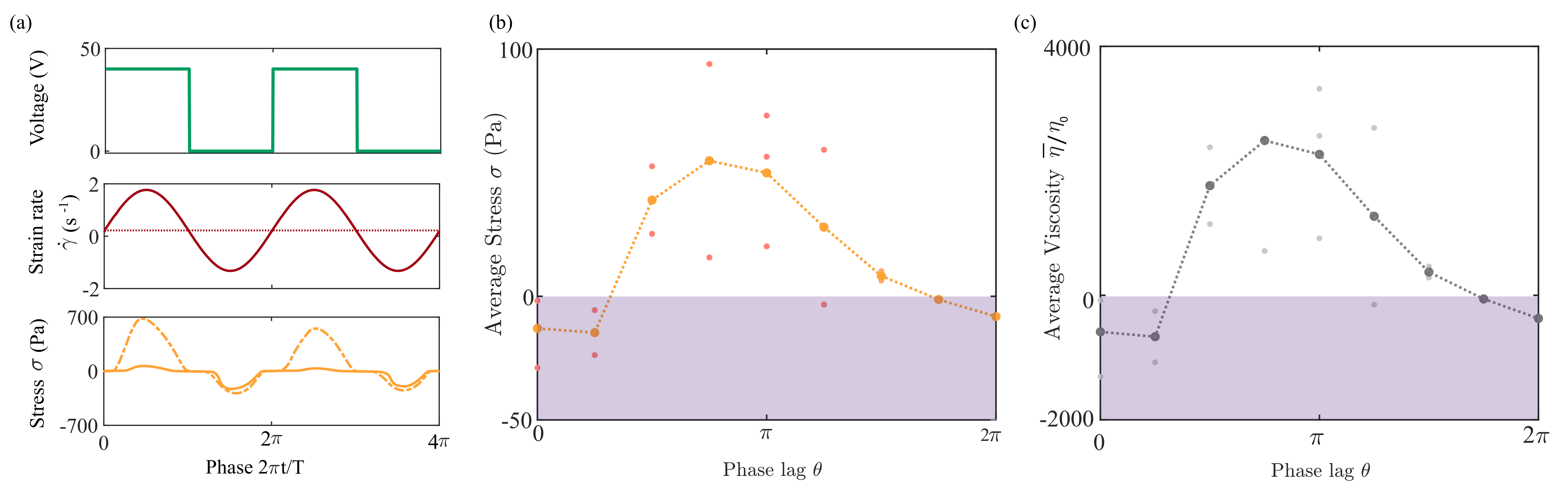}
\caption{\textbf{Negative viscosity states achieved by applying a sinusoidal strain rate} (a) The applied voltage (top), strain rate (middle) and resultant stress (bottom). An oscillatory strain rate is applied with a positive bias. The average of the strain rate (dashed line) is positive. The bottom panel indicates the stress response with (solid line) and without (dashed line) the applied acoustics (b) Time averaged stress for various phase lags between the strain rate and the applied acoustics. The smaller dots are results from individual experiments, whose mean in given by the large connected dots. We find regions of negative viscosity at phase lags close to $2\pi$. (c) Time averaged viscosities normalized by the solvent viscosity for various phase lags. The time averaged viscosity is negative at phase lags close to $2\pi$.} \label{fig:NegativeViscSR}
\end{figure}

We note that negative viscosity states similar to those in Fig~4c can be generated by oscillating the strain rate rather than the stress. Importantly, to generate negative viscosity states while driving the system with an oscillatory strain rate, the acoustics should be applied when the strain rate is positive, reducing the effective stress in the forward direction. Sample voltage, strain rate and stress measurements, as well as the average stress, and viscosity at various phase lags are shown in Fig.~\ref{fig:NegativeViscSR}. We find that the data in strain rate controlled experiments are a lot noisier, both because the rheometer, and the shear thickening process are stress controlled. Despite the noisiness of the data however, we find clear values of the phase lag where the viscosity is negative.  Interestingly, using this protocol, we can also generate states with an effective zero viscosity, where we observe that suspension flows without a net stress.

\clearpage 

\bibliography{pnas-sample.bib}